# Artificial Intelligence in Open Radio Access Network

Paul H. Masur and Jeffery H. Reed, *Fellow, IEEE*

## INTRODUCTION

Fifth-generation (5G, sometimes referred to as 5G New Radio (NR)) wireless networks represent new capabilities for system developers, namely, the potential to tailor systems with end-to-end latency of 1 ms for ultra-reliable low latency communications (uRLLC applications), 1 million terminals per square kilometer for massive machine-type communications (mMTC applications), and a per-user data rate exceeding 50 Mbps for enhanced mobile broadband (eMBB applications) [1]. These capabilities will play an essential role in designing future industrial, aeronautic, and consumer systems. However, to fully leverage 5G, communications networks need to be customized within the overall system from the start. This tailored approach relies upon generic hardware with specialized software to tailor the network to an application's needs and the use of artificial intelligence.

## I. O-RAN BACKGROUND

### A. *O-RAN Origins*

To meet these challenges and opportunities, the O-RAN Alliance was formed in February 2018. The O-RAN Alliance resulted from the merger of the cloud-RAN (C-RAN) Alliance and the extensible-RAN (xRAN) Forum. The groups found similar interest in harnessing open interfaces, Artificial Intelligence/Machine Learning (AI/ML) algorithms, and virtualization as means to meet the 5G International Telecommunication Union (ITU) requirements. Given these overlapping goals, the O-RAN Alliance formally started in June 2018, with the first Work Group meetings occurring in September 2019 [2].

### B. *O-RAN Advantages*

The overarching goals of O-RAN are twofold: 1) integrate proven, cloud-scale economics to RAN, and 2) provide newfound agility to RAN deployments through AI/ML algorithms [3]. Open-source software serves as a foundation for O-RAN compliant functions that meet 5G standard requirements [4]. O-RAN maintains an agile architecture, deployment flexibility, and real-time responsiveness based on changing communication parameters [5].

O-RAN specifically aims to reduce the cost of network deployment by using low-cost, White-box hardware for radio components. This White-box hardware (or Commercial-Off-The-Shelf, COTS) utilizes generic hardware components for processing and radio operations, which employs the aforementioned software. In essence, software supplants hardware, where possible, as a cost-savings mechanism. This is in contrast to previous generations of wireless networks (e.g., 3G, 4G) that relied upon specialized hardware, which came at a greater cost [6], [7]. Further discussion of White-box hardware can be found in Section III(B).

In addition, open interfaces allow for multi-vendor deployments on the RAN. To build an expanded ecosystem, an open-source, multi-vendor approach is taken. These open interfaces are designed for information exchange between AI/ML elements of the O-RAN [8]. A mixed operator environment on the RAN fosters an environment for varied network services through sharing spectral resources. Further discussion of O-RAN Architecture, including open-interfaces, can be found in Section III(A).

The use of generic hardware, however, mandates the use of software for network management across various hardware implementations. The O-RAN alliance seeks to achieve its goals through optimizing traffic steering and load balancing via AI/ML algorithms, prioritizing Quality of Experience (QoE) and Quality of Service (QoS) parameters, and optimizing the beamforming and physical resource block (PRB) usage elements of massive Multiple Input Multiple Output (MIMO)

Paul H. Masur and Jeffrey H. Reed are with the Wireless@VT, the Bradley Department of ECE, Virginia Tech, Blacksburg, VA 24061 USA (e-mail: phmasur@vt.edu; reedjh@vt.edu).

operations [6]. These goals address issues ranging from the Physical to Networking Layers. Each goal represents a particular (or modular) AI/ML implementation. O-RAN implements a chained model approach, as each implementation interoperates within the context of providing user services. This will be further discussed in Section IV.

## II. O-RAN Architecture

The network architecture design of O-RAN intentionally includes the use of COTS. Open (i.e., non-proprietary) interfaces and network components defined at a system level allow for the generic hardware envisioned by O-RAN. The network design and network functions of O-RAN utilize acronyms and a structure similar to the 5G NR standard family, but there are some differences. This section seeks to provide an overview of O-RAN Architecture.

### *A. Higher Level View of O-RAN Architecture*

The network layout of O-RAN relies upon a disaggregated base station model, with different parts of the network stack handled by a variety of different logical and physical units. The radio protocol stack processing is split between the multi-Radio Access Technology (RAT) Central Unit (CU), Distributed Unit (DU), and Radio Unit (RU) [9]. As shown in Fig. 1, the multi-RAT CU handles the upper-level radio stack operations, namely Radio Resource Control (RRC), Packet Data Convergence Protocol (PDCP), and Service Data Application Protocol (SDAP). The multi-RAT CU (or O-CU) is responsible of L2 and Network Layer (L3) operations. The DU (or O-DU) operations, which encompass Medium Access Control (MAC) Layer (L2), include MAC, Radio Link Control (RLC), channel coding, scrambling, and modulation control. Lastly, the RU (or O-RU) handles the RF fronthaul (FH) operations (which lie between baseband processing and the radio equipment), including Physical Layer (L1), digital beamforming and Fast Fourier Transform (FFT)/Inverse Fast Fourier Transform (IFFT) processing. Important interfaces within O-RAN include: the E2 interface for passing data between RAN Intelligent Controllers (RICs), the O1 interface to provide management between O-RAN upper and lower elements, and the A1 interface for interaction between the Near-Real-Time (Near-RT) RIC and Non-Real-Time (Near-RT) RIC [8]. In particular, the E2 and A1 interfaces are critical for AI/ML operations, which are further discussed in Section IV.

Additionally, a flexible RF FH allows for the O-DU and RU to exchange responsibility for different elements of the protocol stack. As more functions are handled by the O-RU, FH bandwidth requirements are reduced, but RF complexity increases, function extendibility becomes more complex, and multi-vendor interoperability becomes more difficult. Alternatively, more functions in the O-DU requires more RF FH bandwidth, but with the added benefit of reduced RF component complexity, and easier extendibility and multi-vendor interoperability [10]. Depending on conditions and cost requirements of the RF FH, the O-RU and O-DU exchange capabilities. This flexibility allows for O-RAN to achieve the 1 ms latency of 5G.

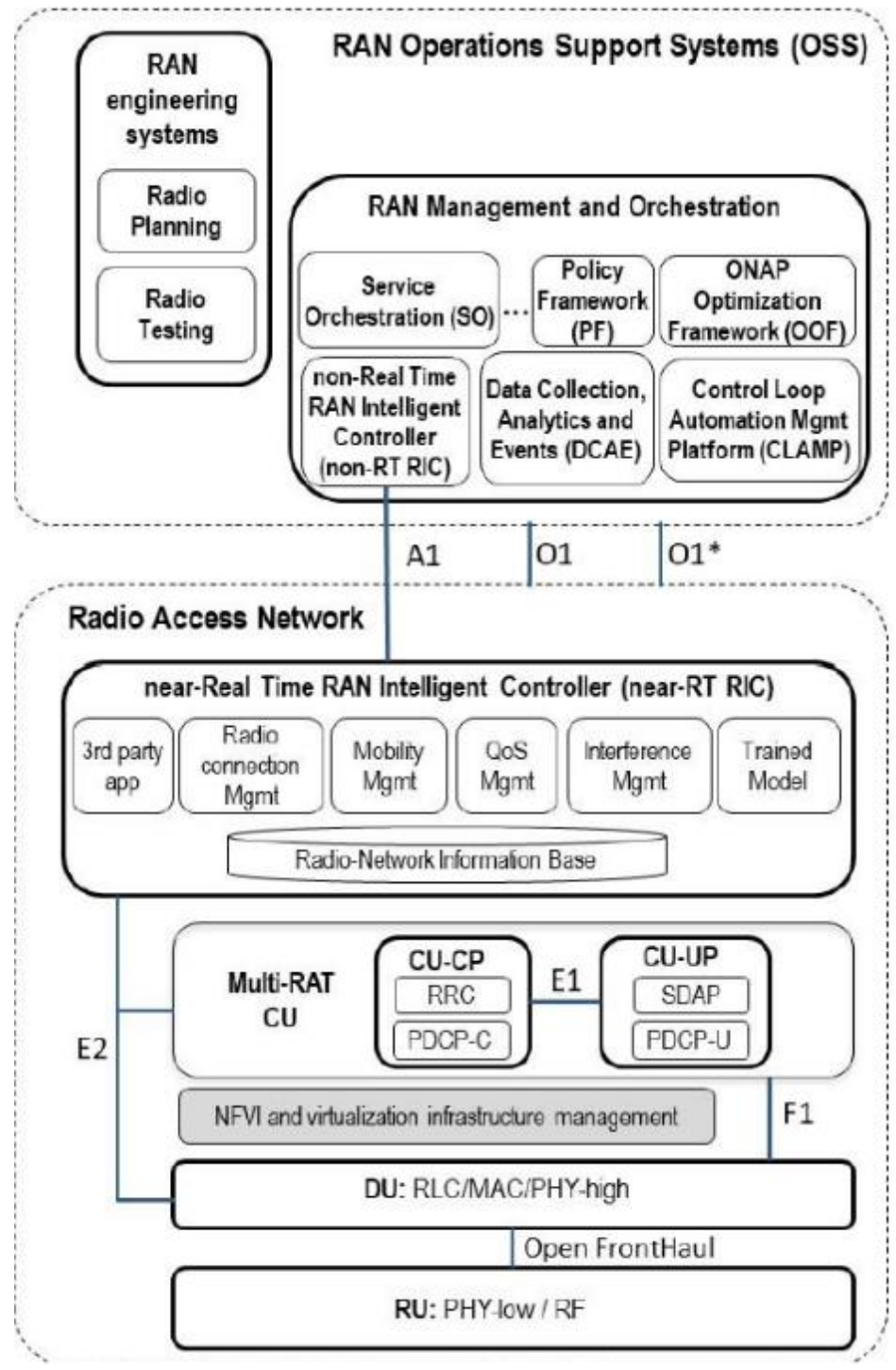

Fig. 1: O-RAN Architecture [9].

### *B. White-box Hardware*

As noted, White-box hardware offers potential for network components within O-RAN to be deployed using generic hardware. The O-RAN Open Fronthaul Interfaces Workgroup (WG4) proposes adopting several split architectures between the O-DU and O-RU to specifically accommodate White-box hardware [11], [12]. O-RAN adopts split architecture options 6, 7-2, and 8, derived from the 8 different FH split architectures scenarios defined by the 3$^{rd}$ Generation Partnership Project (3GPP). Option 6 defines all L1 functions within the O-RU and all L2 functions in the O-DU. Option 7.2, as detailed below, seeks to split L1 between the O-DU and O-RU. Option 8 performs all L1 and L2 functions in the O-DU, thus

Paul H. Masur and Jeffrey H. Reed are with the Wireless@VT, the Bradley Department of ECE, Virginia Tech, Blacksburg, VA 24061 USA (e-mail: phmasur@vt.edu; reedjh@vt.edu).

limiting the O-RU to performing RF to baseband conversion [13].

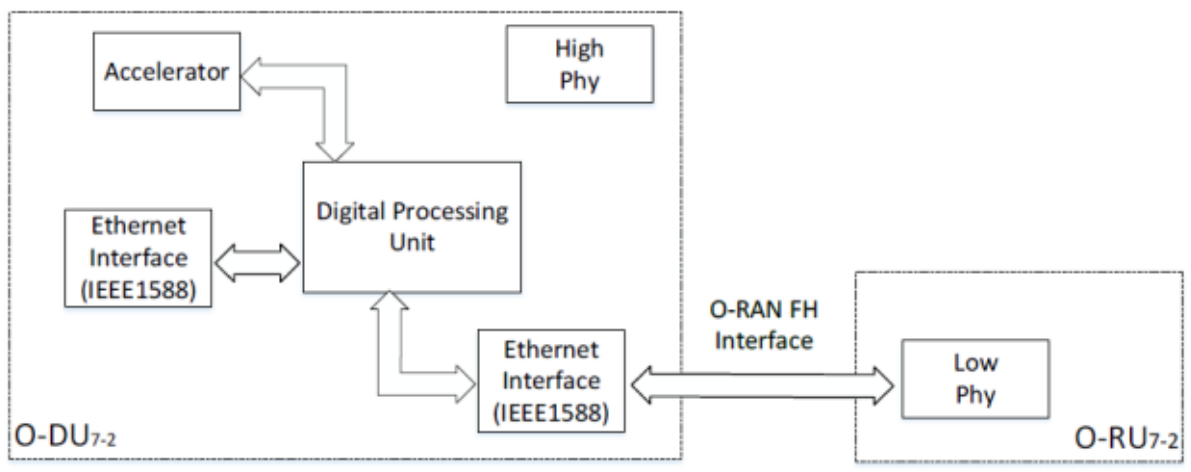

Fig. 2:O-$DU_{7.2}$ with split PHY layer functionality [11].

Fig. 2 displays Option 7.2, representing a balanced separation of functionality between the O-DU and O-RU [12]. This split allows for low Physical Layer operation in the O-RUs and higher physical layer operation in the O-DUs. This is intended to accommodate the 200-300 Gb data transfer required between the O-DU and O-RU, while balancing the processing requirements for handling by generic hardware [12]. The data transfer between the O-DU and O-RU is handled by a fiber-based, RF FH (or O-RAN FH Interface).

As can be seen in Fig. 2, the O-$DU_{7-2}$ allows for generic processing components: a digital processing unit, and a Precision Time Protocol (PTP or IEEE 1588) enabled ethernet interface (for timing with the O-$RU_{7-2}$). The generic hardware is intended to reduce the cost of deployment. As noted previously, O-RAN accommodates a flexible RF FH, which can be integrated into the O-$RU_{7.2}$ directly or be a standalone piece of white box hardware as well [11].

Additionally, O-RAN WG4 lists several key performance indicators (KPIs) for indoor Picocell and outdoor microcell deployment, which O-$DU_{7-2}$ seeks to satisfy. These KPIs include peak data rate, peak spectral efficiency, bandwidth, control plane latency, user plane latency, and QoS at user speed [13]. Within mobile communication network operations, KPIs are used to measure performance and part of QoE/QoS determination. Meeting these KPIs in actual deployment scenarios is critical for achieving 5G use cases.

## III. Machine Learning Principles and Types

While Fourth generation (4G) Long Term Evolution (LTE) deployments originated the use of high-speed, packet-based wireless links, the uplink/downlink (UL/DL) design was a general-purpose radio frame (i.e., single sub-carrier spacing and slot sizes). As a new generation of RAN is designed with varying configurations in the radio frame, engineers seek to rely upon ML techniques to achieve greater efficiencies in RAN design and Physical Layer operation. Moreover, ML techniques allow for a greater variety of use cases within the same spectral resources (e.g., through network slicing). For O-RAN development, these ML techniques are categorized into supervised learning, unsupervised learning, and reinforcement learning. The sections below provide a brief overview of these categories.

### A. *Supervised Learning*

Supervised learning defines several ML techniques that seek to learn a mapping function between a known input data set and a known output data set. A training phase is heavily relied upon to build a generalization capability of the supervised learning technique. The training phase uses known inputs and creates its own labeling algorithm to map the data to the desired output scenario. The ML technique is then tested against a subsequent data set with the same distribution as the original training data set. From this analysis, a training error can be evaluated, which is used to further hone and train the supervised learning algorithm [1].

Supervised learning is used for classification: relating input data sets to output data sets [14]. Supervised learning techniques are expected to be used for QoE optimization, traffic steering, QoE-based traffic steering, and Vehicle-to-Everything (V2X) handover management [15]. One can easily see that V2X handover management relies on many conditions that dictate when a handover should occur, and expediting this process is critical to achieving the 1ms latency requirement of 5G. Moreover, QoE and traffic steering generally operate according to several routing and scheduling rules. Supervised learning is well suited to learn, implement, modify, and eventually optimize these rules [9].

### B. *Unsupervised Learning*

Unsupervised learning uses unstructured input data and provides a relationship between the data set to perform a particular task [1]. Unsupervised learning is used to discover relationships between points within the unlabeled input data set. Unlike supervised learning, there is no uniform theoretical basis underpinning unsupervised learning. Instead, the algorithm seeks to learn by observation without any explicit feedback [14].

While the O-RAN Alliance defines many unsupervised learning techniques in its early standard documentation, the defined uses for this family of ML techniques are for future study (FFS) [15]. However, it has been noted that unsupervised learning shows promise in resource allocation in network design and resource allocation management [1]. Given the scarcity

Paul H. Masur and Jeffrey H. Reed are with the Wireless@VT, the Bradley Department of ECE, Virginia Tech, Blacksburg, VA 24061 USA (e-mail: phmasur@vt.edu; reedjh@vt.edu).

of spectral resources, further optimizations in this domain are highly lucrative.

### C. *Reinforcement Learning*

Reinforcement learning relies upon elements of supervised and unsupervised learning [14]. As with supervised learning, reinforcement learning begins with a known input training data set and creates mapping rules for a known output data set. The goal of reinforcement learning is to optimize long-term conditions [15]. To achieve this end, the algorithm interacts with environmental data, which is theoretically similar to unsupervised learning. Through this environmental interaction, feedback loops are used to periodically update based on changes to the wireless network environment [9]. Allowing for data set changes over time bestows the opportunity to modify and optimize operation [1].

The family of reinforcement learning techniques shows much promise for future network implementation. Since wireless networks already have metrics to define clear outcomes, e.g., channel conditions, signal strength, throughput, latency, QoE, QoS, the feedback loop of reinforcement learning offers an untapped opportunity to tweak network settings to achieve optimal output [14]. Moreover, given the transient nature of spectral usage over time and geography, ML techniques extend the opportunity for enhanced network planning.

## IV. AI Applied to O-RAN Deployment

ML techniques are best suited to handle the vast number of parameters and options that apply to planning and operating a wireless communications network. Indeed, the likely early deployment of ML techniques will combine supervised, unsupervised, and reinforcement learning algorithms to expedite the identification and rectification of network issues [16]. Applying ML to improve channel estimation/detection, channel coding, modulation, layer mapping, antenna pre-coding, OFDM signal generation, and beamforming has been the focus of previous research efforts. Further, ML techniques have been found to improve MAC operations, including dynamic resource allocation, link adaptation, beam management, and hybrid automatic repeat request (HARQ) [9], [14], [17]. An important topic in this area is balancing spectral efficiency and fairness to specific network services/users [18]. As ML techniques are further introduced, further optimization in these areas is expected.

The AI/ML applications within O-RAN are applied according to four overarching principles. First, offline learning must occur before deployment within the network, as most current algorithms need a certain level of seeding before deployment. Second, AI/ML models are trained through a combination of supervised, unsupervised, and reinforcement learning. Third, AI/ML algorithms are implemented in a modular manner to allow for independence. Lastly, O-RAN and its open interfaces allow for vendor-specific AI/ML algorithms within each module [15]. While the AI/ML algorithms have independence, the A1 and E2 interfaces and parallel operation of xApps (further discussed in Section V(B)) allow cross-layer interaction. While this cross-layer interaction may offer stability through redundancy checking, optimizing the interaction offers additional challenges.

AI/ML incorporation into the RAN itself is not without challenges. While Physical layer and MAC operations are distinct within the logical OSI model, there is a significant interplay between these operations, and one’s performance affects the other. Communication and signaling on the E2 and A1 interfaces are essential, as these interfaces tie together the training loops (See Fig. 3). Therefore, orchestration between these techniques is critical for proper deployment. Proper coordination, updating, and training to the AI/ML algorithms are essential. O-RAN must handle cross-layer communication of data for AI/ML while protecting previous cross-layer communications [5]. The cross-layer communications and separation of the AI/ML elements present a harmonization challenge [19].

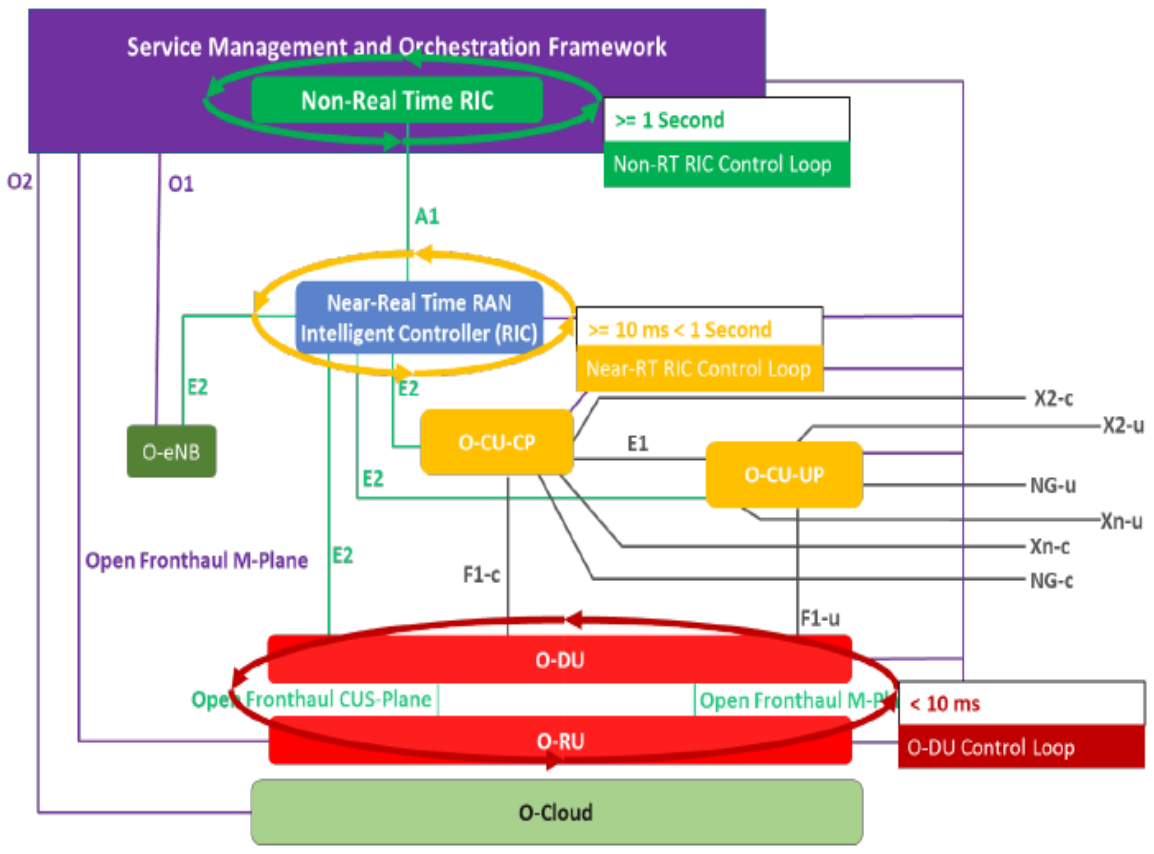

Fig. 3: AI Control Loops within O-RAN [8].

### A. *ML Modeling within O-RAN*

The O-RAN specification currently defines different types of ML model design. These include chained modeling, singular modeling, and chained modeling with Paul H. Masur and Jeffrey H. Reed are with the Wireless@VT, the Bradley Department of ECE, Virginia Tech, Blacksburg, VA 24061 USA (e-mail: phmasur@vt.edu; reedjh@vt.edu).

common inputs. Chained modeling allows modular ML models to be separately implemented and chained together to create a single output. Singular modeling is a standalone implementation allowing for many inputs to an ML algorithm with a single output. A chained model with common inputs allows later stages of the chain to access earlier stage outputs and inputs [15].

The chained model approach allows for the reuse of existing models and chaining to provide a single determination, e.g., QoE determination. Furthermore, an interference host can communicate between separated yet parallel models [15]. This implementation allows for xApp deployment and for coordinating between xApps to make simultaneous determinations at the non-RT RIC and near-RT RIC (see the section below).

### B. *Non-Real-Time (Non-RT) RIC*

Non-RT data is readily available from existing network management systems, which provides an opportunity to model and simulate existing planning algorithms [18]. As shown in Fig. 3 and Table I, the control loop timing of Non-RT RIC is greater than one second, and in some instances, much, much greater. Table I shows the timescale differences between the AI Control Loops within O-RAN. The current focus of NRT RIC applications includes cell planning, frequency reuse, and meeting capacity based on time of day. These improvements have been long sought after and show great promise, particularly at predicting and mitigating congestion [20]. The Non-RT RIC yields an opportunity for long-term network planning and optimization.

TABLE I. A COMPARISON OF AI CONTROL LOOPS IN O-RAN

| Control Loop | Timescale | Area of Influence |
|---|---|---|
| Non-RT RIC | $\geq$ 1s | Supervisory function, spectral use, and cell/network planning |
| Near-RT RIC | 10ms $\leq$ x $\leq$ 1s | Radio Frame Setting, L2/L3 Operations, PBCH Setting |
| O-DU Control Loop | $\leq$ 10 ms | FFS – NR Numerology (or frame structure) setting, L1/L2 Operations, Synchronization |

Since the non-RT RIC operates on a longer time duration (greater than 1 s) than the near-RT RIC, supervisory functionality is also supported. Furthermore, the non-RT RIC orchestrates data collection and drives data inputs for the near-RT RIC. This process provides a more efficient operation of the near-RT RIC [21]. The communication between the Non-RT RIC and Near-RT RIC occurs over the A1 interface, as shown in Fig. 3.

### C. *Near-Real-Time (Near-RT) RIC*

Near-RT RIC shows an opportunity for handling user plane and control plane management. The control loop operates on a 10 ms interval [8]. Current techniques in physical layer handling and MAC scheduling are the results of traditional design iterations. While these human-led techniques have been honed and refined over decades, ML promises improvement on these techniques and newfound adaptability. A timeframe of 10 ms to 1 s provides a level of agility for the network to operate. However, data collection for Near-RT RIC needs to be obtained from the E2 interface through existing user equipment (UE), cell, and network parameter measurement [18]. Therefore, the Near-RT is dependent on other functions to meet its operational requirements.

Given the timescale shown in Table I, the near-RT would be expected to handle operations at or above the timescale of a 5G NR radio frame (which has a 10 ms length). Information disseminated in the Physical Broadcast Channel (PBCH, sent every 40ms), including the Master Information Block (MIB) could be tuned according to the near-RT RIC Controller. Additionally, slow fading can be handled on this timescale using L2 operations like link adaptation and beam management.

The Near-RT RIC contains several xApps (or different applications) running on Application Programmable Interfaces (APIs), shown in Fig. 4. These xApps may run in parallel to one another. The xApps further handle many management and/or ML functions, as they may connect via the A1 interface and/or E2 interface. Conflict mitigation and subscription management are used to coordinate and manage xApp outputs. This functionality must coordinate between xApps for timing demands, monitor for proper configuration, balance processing resources, and handle messaging [22]. As noted previously, this is a logical implementation of the interference host for ML modeling configurations.

The use of xApps allows for the non-RT RIC to take a modular approach to ML algorithms. In essence, each xApp assumes responsibility for a different ML algorithm application. The non-RT RIC or near-RT RIC can then perform hosting operations for the xApps – to either collect data for further use or implement changes in network operation. The use of Deep Reinforcement

Paul H. Masur and Jeffrey H. Reed are with the Wireless@VT, the Bradley Department of ECE, Virginia Tech, Blacksburg, VA 24061 USA (e-mail: phmasur@vt.edu; reedjh@vt.edu).

Learning (DRL) as an xApp has been shown to yield 20% gains in both spectral efficiency and buffer occupancy [23].

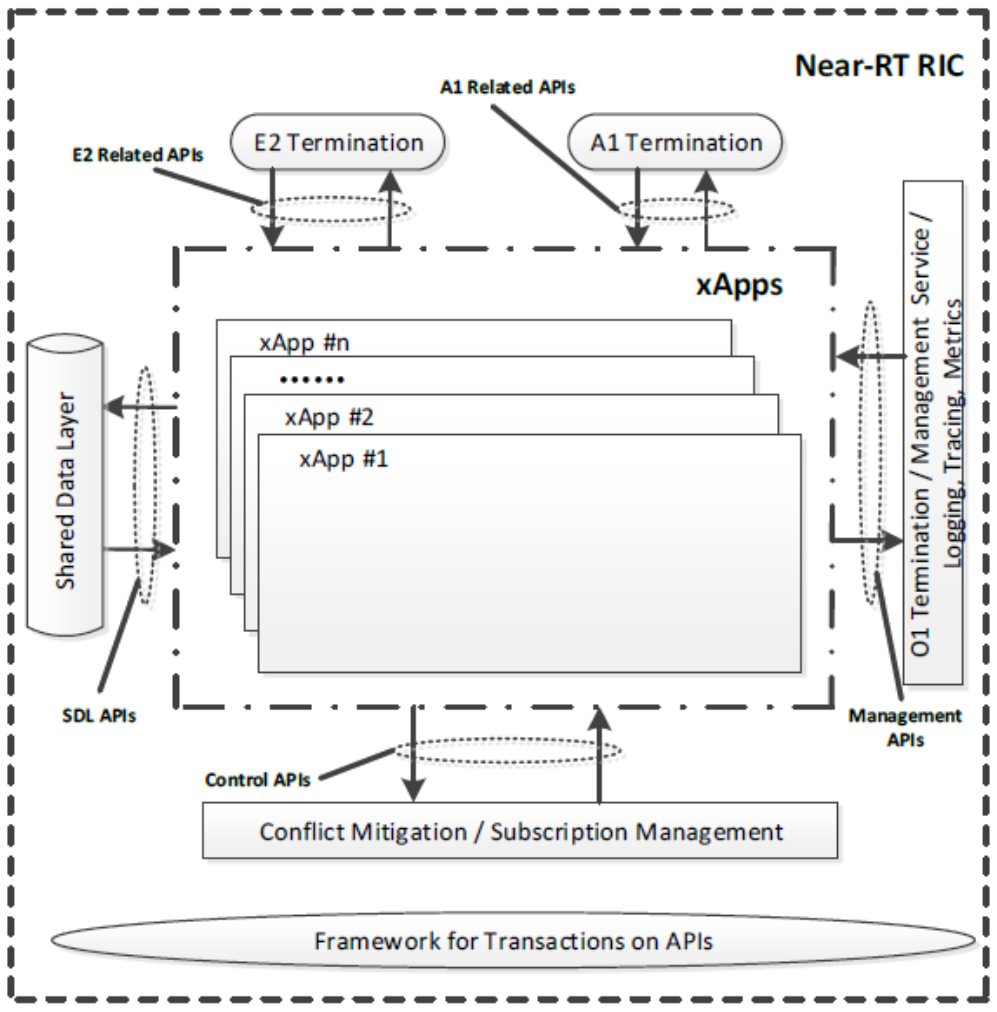

Fig. 4. Detailed view of near-RT RIC [22].

## *D. O-DU and O-RU AI Control Loop*

The control loop operations at the O-DU and O-RU are largely considered to be FFS. This operation occurs on a transmission time interval (TTI) basis, so on the order of 10 ms or less [15]. As shown in Table I, this operates on the smallest timescale envisioned by O-RAN. Given the length of a 5G NR radio frame, operations could occur on the frame, sub-frame, or even PRB level.

On this small-time scale, channel estimation and PRB usage offer potential areas for future ML applications. Fast fading, where the channel impulse response changes faster than symbol duration, is another potential application. An ML application with prior access to channel conditions (e.g., a Radio Environment Map (REM)) could offer a proactive response to fast fading at the UE and O-DU. L2 operations, including synchronization and 5G numerology (or frame structure) setting, are likely areas to be influenced by the O-DU loop. 5G NR relies more heavily on TDD spectrum, so UL/DL slot setting is crucial for achieving latency and bandwidth requirements. Proper use of synchronization signals enables UL/DL setting, and the O-DU loop could play a greater role in this regard. Additionally, the numerology setting of a 5G NR radio frame affects slot length (and thus latency and bandwidth), so effective, timely tuning of this setting is also important. Lastly, measurement signal setting and reporting fall under L1 operations, which would allow for expedited radio parameter setting.

While the O-DU and O-CU are seen as the elements of a real-time control loop, the UE could also perform operations [23]. An astute observer might notice that O-RAN outlines functionality and requirements for RAN but is silent regarding the UE. Traditionally, functionality is added to the RAN before the UE, because the RAN design is not as susceptible to power, size, and component cost requirements as UEs. However, given round trip time (RTT), eventually harnessing the UE for some AI Control Loop decision making through the E2 interface makes sense. Moving a UE beyond a data reporter/collector could further reduce latency in decision-making. For applications like V2X, any time saved translates into saved lives. The developing field of Cognitive Radios (CRs) offers potential for realizing the UE's future inclusion into the AI Control Loops of O-RAN.

The introduction of the UE to the AI Control Loops begs some interesting questions. Could this be an area to identify and rectify self-interference at the UE? Use cases in 5G mandate a 1ms latency and high accuracy, so retransmission (HARQ) due to self-interference in a full-duplex model should be avoided wherever possible. Each radio subframe within 5G contains several reference and measurement signals (e.g., the Channel State Information Reference Signals, CSI-RS). The O-DU and O-RU may adjust modulation, power, and beamforming to assist with self-interference identification and mitigation at the UE.

## *E. ML and Security: Solutions and Issues*

The use cases of 5G, from self-driving vehicles to smart power grids, make security a paramount concern. Moreover, the use and availability of hardware devices, whether COTS or specialized, come with their own security concerns. The supply chain of telecommunications equipment raises many security concerns [24]. O-RAN is seen as an effort to counteract these concerns, as it offers the ability to meet 5G requirements while using as much generic hardware as possible. A concern has been raised about whether an open-sourced internet sponsored by Western countries and a closed-source internet implemented by others may bifurcate the internet entirely, erasing many of the openness gains that have been achieved through the years [24]. That said, the supply chain security concerns have no easy solution.

Devices within O-RAN present the ability to run software that does not trust its underlying hardware. The Paul H. Masur and Jeffrey H. Reed are with the Wireless@VT, the Bradley Department of ECE, Virginia Tech, Blacksburg, VA 24061 USA (e-mail: phmasur@vt.edu; reedjh@vt.edu).

hardware is validated through hardware signaling checks and software validation [25]. Additionally, the open-source software aspects from O-RAN allow for defensive (i.e., encryption) and offensive (i.e., proactively securing resources when faced with jamming) capabilities at devices [17]. Open-source software presents security challenges regarding its open nature but has the advantage of rigorous testing, varied testing, and customization against threats.

Adversarial learning is an area that applies to ML and security within O-RAN. Adversarial learning develops attacks to thwart or modify existing ML algorithms without detection. One could easily see how targeted attacks could result in denial-of-service (DoS) within the O-RAN network. One such vulnerable area includes CSI feedback parameters used for communication parameter setting and allocation of resources [26]. Additionally, certain Adversarial attacks include synthetic data to fool existing traffic steering ML models [27]. Jamming attacks are another avenue of adversarial ML, where an attack is only initiated if an acknowledgment (ACK) is predicted [28]. Lastly, Trojan horse attacks can be deployed on modulation symbol classification by altering training data, thus degrading communication ability [29].

The above attacks introduce vulnerability to the non-RT RIC (for traffic steering models) and the near-RT RIC (for CSI feedback modification) operations within O-RAN. Therefore, as the O-RAN standard is developed, it is incumbent upon designers to anticipate these issues. As previously noted, CR and more advanced UEs may provide some potential solutions in the future. The ability to crowdsource data may allow for the network to identify and mitigate bad actors.

Further, future works should answer these critical questions:

- Can a network run on zero-trust hardware?
- Can algorithms be developed to find illicit monitoring? Can counter intelligence be performed to modify illicit monitoring results, effectively employing digital counterintelligence? This need appears to be a future application of adversarial learning.
- Can countries employ their own monitoring without being subject to counterintelligence? There are two sides to this coin.

## V. Conclusion

The future of wireless network design lies firmly with AI/ML techniques [30]. O-RAN offers a promising future for wireless communications, as ML techniques are built directly into the operation of the network architecture itself. The direct integration of these ML techniques allows for their presence to be optimized within the network instead of layering them on top of an existing architecture. Furthermore, the allowance of White-box hardware and open-source software lowers the entry cost into the market. Given the vast possibility of consumer offerings through the use of network slices, O-RAN serves as a potential backbone for the type of fast yet agile network deployment that is needed to meet or exceed the lofty use case requirements of 5G.

Past network design relied on human processes: data collection, modeling, and testing. Future implementations are poised to harness powerful learning techniques (both supervised and unsupervised) to produce the network layouts that will foster network services. Simply put, ML yields a faster and more accurate method of optimizing RAN layout to individual communication parameters. As noted earlier, O-RAN's use of AI/ML offers an attractive business opportunity to provide a multi-vendor environment, plus unique properties for improved system designs.

## VI. *Acknowledgment*

The authors would like to thank Tarun Cousik and Dr. Vijay Shah of Wireless@VT for their constructive feedback. The authors additionally acknowledge Dr. Brian Agee for his comments and suggestions.

Paul H. Masur and Jeffrey H. Reed are with the Wireless@VT, the Bradley Department of ECE, Virginia Tech, Blacksburg, VA 24061 USA (e-mail: phmasur@vt.edu; reedjh@vt.edu).

Paul H. Masur and Jeffrey H. Reed are with the Wireless@VT, the Bradley Department of ECE, Virginia Tech, Blacksburg, VA 24061 USA (e-mail: phmasur@vt.edu; reedjh@vt.edu).

Paul H. Masur and Jeffrey H. Reed are with the Wireless@VT, the Bradley Department of ECE, Virginia Tech, Blacksburg, VA 24061 USA (e-mail: phmasur@vt.edu; reedjh@vt.edu).